\documentclass[twocolumn,10pt,prl,showpacs,preprintnumbers]{revtex4-1}%

 \usepackage{graphicx}
\usepackage{epsfig}

\usepackage{hyperref}
\hypersetup{
     colorlinks=true,
    linkcolor=blue,          
    citecolor=blue,        
    filecolor=blue,      
    urlcolor=blue          
}

\usepackage{verbatim}
\usepackage{amsmath}
\usepackage{amsfonts}

\begin{document}

\preprint{UUITP-03/12, DCPT-18/35}
\title {\Large\bf Meta-stable non-extremal anti-branes}
 \author{\bf J. Armas$^{a, b}$, N. Nguyen$^c$, V. Niarchos$^c$, N. A. Obers$^d$, T. Van Riet$^{e}$\\}
\vspace{0.8cm}
\affiliation{$^{a}$ Institute for Theoretical Physics, UVA, 1090 GL Amsterdam, The Netherlands}
\affiliation{$^{b}$ Dutch Institute for Emergent Phenomena, The Netherlands}
\affiliation{$^{c}$Department of Mathematical Sciences and Centre for Particle Theory\\ Durham University, Durham DH1 3LE, UK}
\affiliation{$^{d}$ The Niels Bohr Institute, University of Copenhagen\\ Blegdamsvej 17, DK-2100 Copenhagen $\O$, Denmark}
\affiliation{$^{e}$Instituut voor Theoretische Fysica, K.U.Leuven,\\Celestijnenlaan 200D, B-3001 Leuven, Belgium}
\printfigures

\begin{abstract} 
We find new and compelling evidence for the meta-stability of SUSY-breaking states in holographic backgrounds whose consistency has been the source of ongoing disagreements in the literature. As a concrete example, we analyse anti-D3 branes at the tip of the Klebanov-Strassler (KS) throat. Using the blackfold formalism we examine how temperature affects the conjectured meta-stable state and determine whether and how the existing extremal results generalize when going beyond extremality. In the extremal limit we exactly recover the results of Kachru, Pearson and Verlinde (KPV), in a regime of parameter space that was previously inaccesible. Away from extremality we uncover a meta-stable black NS5 state that disappears near a geometric transition where black anti-D3 branes and black NS5 branes become indistinguishable. This is remarkably consistent with complementary earlier results based on the analysis of regularity conditions of backreacted solutions. We therefore provide highly non-trivial evidence for the meta-stability of anti-branes in non-compact throat geometries since we find a consistent picture over different regimes in parameter space.   
\end{abstract}


\maketitle

\paragraph{Introduction.}
An understanding of controlled supersymmetry (SUSY) breaking in string theory is arguably one of the most important goals in order to progress string phenomenology. Thanks to holography, the use of controlled SUSY breaking in string backgrounds would improve our understanding of strongly coupled phenomena in field theories with broken SUSY. The latter is an important issue independent of the conjecture that string theory is the quantum theory underlying our physical world.  

One of the canonical methods for SUSY breaking in string theory employs anti-branes in warped throats. The study of this mechanism has a long history, starting with the original papers \cite{Maldacena:2001pb, Kachru:2002gs} whose motivation was the holographic description of dynamical breaking of supersymmetry (SUSY). Since then, anti-branes have become an indispensable tool to break SUSY in various contexts. The applications beyond holography include de Sitter \cite{Kachru:2003aw}, inflationary model building \cite{Kachru:2003sx} and the construction of non-extremal black hole micro-states \cite{Bena:2012zi}. 
For some of these applications additional complications due to the compactness of extra dimensions arise, which we will not address. Instead, we restrict to the study of anti-branes in the non-compact KS throat. The motivation for that is two-fold. On one hand, the KS throat is a prime example of a top-down holographic background with broken conformal symmetries that exhibits confinement. Studying SUSY-breaking in this context is therefore a concrete example of the ultimate goal to get a computational handle on QCD-like field theories in the absence of SUSY. On the other hand, for string phenomenology it has been argued that the KS throat at least constitutes a local description of the more elusive compact throat geometries. 

In this letter we present entirely novel arguments for the highly debated meta-stability of SUSY-breaking states in these throats and, at the same time, demonstrate how to extend the analysis to situations that incorporate finite temperature effects in the holographic dual, offering new methods to study phase transitions in this context. 

Our candidate meta-stable states are described by $\overline{\rm D3}$'s at the tip of the throat. They can decay because the surrounding 3-form fluxes induce delocalised D3 charges  out of which D3's can nucleate and annihilate with the $\overline{\rm D3}$'s. This process can be described in terms of brane polarisation in which the $\overline{\rm D3}$'s ``puff" into a spherical NS5  wrapping a contractible $S^2$ inside the $S^3$-cycle \cite{Kachru:2002gs}. As the NS5 moves over the $S^3$, it changes the sign of the D3 charges it carries, effectively mediating the brane-flux decay.  To find a meta-stable state the NS5 needs to find a balance between the $H_3$-flux force that wants to push the NS5 over the $S^3$ and the force of its own ``weight" doing the opposite. In the probe limit, KPV found that such a balance of forces exists  whenever the ratio $p/M$ is small enough \cite{Kachru:2002gs}. Here $p$ denotes the number of $\overline{\rm D3}$'s and  $M$ the quantum of 3-form flux  piercing the $S^3$.

The existence of this KPV state has been refuted in various works starting with the investigations of \cite{Bena:2009xk}. The problem, found at the time, arises when trying to go beyond the probe limit and investigate what happens once the branes backreact. In particular \cite{Bena:2009xk}, and many subsequent works \cite{Gautason:2013zw}, found that the backreacted geometry had singular 3-form fluxes in such a way that it would cause immediate brane-flux decay \cite{Blaback:2012nf}. As a response, \cite{Michel:2014lva} argued that the singularity is renormalised in such a way that does not affect stability when $p=1$, which is a case that is not amenable to a supergravity analysis. Refs.~\cite{Cohen-Maldonado:2015ssa, Cohen-Maldonado:2016cjh}  argued that meta-stability can also be retained when $p\gg 1$ since the observed singularities cannot be proven to exist once one backreacts spherical NS5's instead of point-like $\overline{\rm D3}$'s. In fact, all proofs of unphysical singularities rested on an assumption which was in contradiction with KPV from the start, since $\overline{\rm D3}$ meta-stable states are really NS5 states. 

Several studies \cite{Bena:2012ek,Bena:2013hr, Blaback:2014tfa, Hartnett:2015oda} have investigated the effect of adding temperature to the $\overline{\rm D3}$'s. Most of these works were motivated by the would-be singularity in the 3-form fluxes. Whether or not a singularity can be cloaked by a horizon that arises when moving away from extremality is believed to be an important criterium for deciding the fate of singularities \cite{Gubser:2000nd}. Although strong indications were found that one should not worry about singularities at all, \cite{Michel:2014lva, Cohen-Maldonado:2015ssa,Cohen-Maldonado:2016cjh}, it remains an outstanding problem to understand what happens when the $\overline{\rm D3}$'s are at finite temperature.
If the state would destabilise infinitesimally away from extremality, it would be a sign of being gapless, which is not wanted. In this letter we present additional evidence that this does not happen and provide new, previously inaccessible, quantitative results that support the picture of \cite{Cohen-Maldonado:2015ssa, Cohen-Maldonado:2016cjh}.

\vspace{0.2cm}
\paragraph{Our approach.}

\vspace{-0.2cm}
From the NS5 viewpoint D3-NS5 polarisation is most naturally described in the supergravity regime where $g_s p \gg 1$ (see e.g.\ \cite{Bena:2014jaa}). This involves a daunting task: solving the type IIB supergravity equations to find D3-NS5 bound state solutions wrapping an $S^2$ in the presence of the fluxes of the KS background. In this paper, we attack this problem using blackfold techniques \cite{Emparan:2009cs, Emparan:2009at, Armas:2016mes}. 

In the blackfold formalism, problems of the above type are treated systematically by setting up a scheme of matched asymptotic expansions, where the solution is approximated in a far-zone by the background solution of interest (here the KS background) and in a near-zone by a uniform flat-space $p$-brane solution (here the D3-NS5 bound state). This scheme is possible when the characteristic length scales of the near-zone solution, denoted collectively by $r_b$, are hierarchically smaller than the characteristic length scales $\mathcal R$ of worldvolume inhomogeneities, and the characteristic length scales $L$ of the background. The regime $r_b \ll \mathcal R,L$ is a regime of long-wavelength expansions. For a general discussion of blackfolds and the approximations they entail we refer the reader to \cite{Armas:2016mes} and references therein. The specifics of the regimes of interest in our context are summarised below. 

A key component of the above analysis reformulates part of the supergravity equations as an effective worldvolume theory. In the case at hand, this is a supergravity-derived 6d effective theory that describes the long-wavelength properties of NS5 branes with dissolved 3-brane charge. In what follows, we focus exclusively on the leading-order equations of the 6d theory ({\it blackfold equations}). At the very least, these equations pose necessary conditions for the existence of the long-sought supergravity solution in the above regime.

\begin{figure*}[!]
\begin{center}
\includegraphics[width=0.45\textwidth]{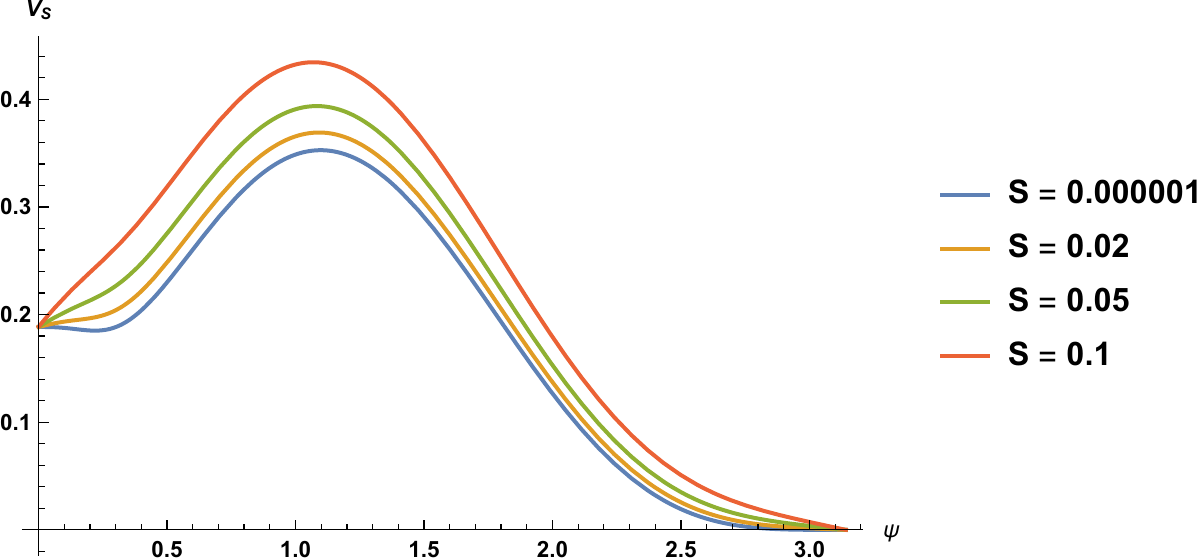} \hspace{1.5cm}
\includegraphics[height=3.8cm,width=0.35\textwidth]{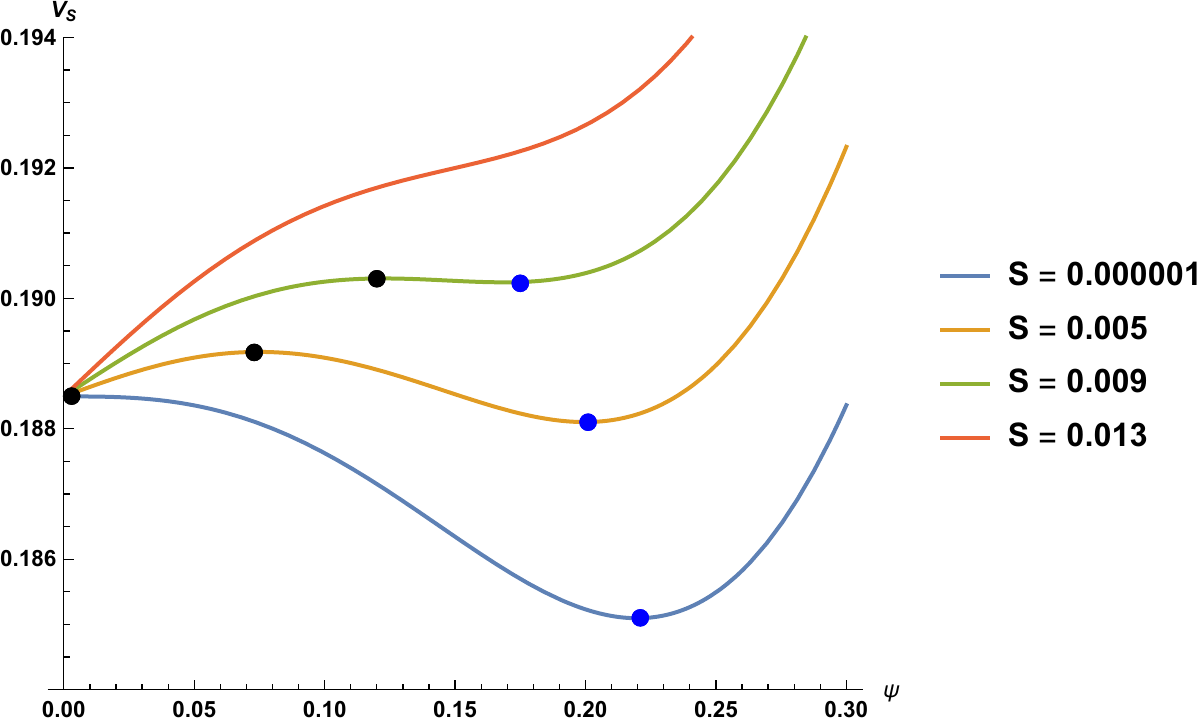}
\end{center}
\vspace{-0.3cm}
\caption{Plots of the effective potential at fixed entropy, $V_S$, as a function of the angle $\psi$ on $S^3$. Both figures represent plots at $p/M=0.03$. The right plot zooms into the region near the North pole of the $S^3$. 
As we increase the entropy we encounter a critical value $S^*$, where the meta-stable vacuum of KPV (blue dots on the right) merges with a new unstable vacuum (black dots on the left). 
}
\vspace{-0.3cm}
\label{fig1}
\end{figure*}

\vspace{0.2cm}
\paragraph{Forced blackfold equations.}

The leading order blackfold equations for black branes in the presence of general external fluxes in (super)gravity were obtained in \cite{Armas:2016mes}. We consider ${\overline {\rm D3}}$-NS5 branes at the tip of the KS throat, where in appropriate units the string frame metric is 
\cite{Klebanov:2000hb} 
\vspace{-0.5cm}
\begin{equation}
\label{geneqaa}
ds^2 = g_s M b_0^2 \ell_s^2 \left( ds^2({\mathbb M}_{0123}) + d\Omega_3^2 + dr^2 + r^2 d\Omega_2^2 \right).
\end{equation} 
Here, ${\mathbb M}_{0123}$ is Minkowski space in the directions 0123, $d\Omega_n^2$ is the metric element of the round unit $n$-sphere, $b_0^2 \simeq 0.93266$ and $\ell_s$ the string scale. There is also a RR field strength $F_3=dC_2 = 2M \ell_s^3 {\rm Vol}(S^3)$ across the $S^3$ and a 7-form flux $H_7=dB_6$ dual to $H_3=dB_2$
\vspace{-0.2cm}
\begin{equation}
\label{geneqab}
H_7= \frac{1}{g_s^2} \star H_3 = - \frac{\ell_s^4}{g_s} dx^0 \wedge dx^1 \wedge dx^2 \wedge dx^3 \wedge F_3
~.\vspace{-0.2cm}
\end{equation}
The dilaton is constant and the self-dual 5-form field strength vanishes at the tip.

The boosted D3-NS5 bound state solution in flat space, which forms the seed of our blackfold expansion, is well known
(see e.g. \cite{Alishahiha:1999ci,Harmark:1999rb}). 
 The following thermodynamic data of this solution (in the Einstein frame) are needed for our purposes: the energy-momentum tensor
\begin{eqnarray}
\begin{split}
\label{geneqac}
T_{ab} &= {\mathcal C} \bigg[ r_0^2 \left( u_a u_b -\frac{1}{2}\gamma_{ab} \right) - r_0^2 \sin^2 \theta \, \sinh^2 \alpha \, \hat h_{ab} 
\\
&- r_0^2 \cos^2\theta \, \sinh^2\alpha \, \gamma_{ab} \bigg]
~,
\end{split}
\end{eqnarray}
\vspace{-0.05cm}
and the charge currents
\vspace{-0.15cm}
\begin{eqnarray}
\label{geneqad}
J_2 &=& \mathcal C r_0^2 \sinh^2\alpha \, \sin\theta \cos\theta\, v \wedge w~,
\\
J_4 &=& \mathcal C r_0^2 \sinh\alpha \cosh\alpha \sin\theta \, * (v\wedge w)~,
\\
\mathfrak j_6 &=& \mathcal C r_0^2 \sinh\alpha \cosh\alpha \cos\theta \, * 1
~,
\end{eqnarray}
where $\mathcal C=\frac{1}{(2\pi)^5 g_s^2 \ell_s^8}$ and $\gamma_{ab}$ is the induced metric on the fivebrane worldvolume (parameterised by $\sigma^a$ with $a,b,\ldots=0,1,\ldots,5$). The Hodge dual of $\gamma_{ab}$ is $*$ and $\hat h_{ab}$ is a projector onto the directions of the dissolved $\overline{\rm D3}$-brane charge inside the fivebrane. General boosts/rotations of the D3-NS5 bound state solution are expressed in terms of the velocity timelike unit vector $u^a$ and the spacelike orthonormal vectors $v^a, w^a$. In terms of these vectors $\hat h_{ab}= \gamma_{ab}-v_a v_b - w_a w_b$. The electric currents $J_4$, $\mathfrak j_6$ express the D3, NS5 currents of the solution while $J_2$ is a consequence of the non-zero $C_2$-field of the solution. $r_0$ is the Schwarzschild radius. In the extremal limit, $r_0 \to 0$, $\alpha\to \infty$ while the combination $r_0^2 e^{2\alpha}$ is kept fixed. The parameter $\theta$ controls how much $\overline{\rm D3}$ brane charge is dissolved inside the NS5 brane.

The general effective blackfold equations of the $\overline{\rm D3}$-NS5 brane in the presence of 3-form NSNS/RR fluxes, constant dilaton and vanishing 1-, 5-form RR fluxes are \cite{Armas:2016mes}
\vspace{-0.2cm}
\begin{eqnarray}
\label{geneqae1}
\nabla_a T^{a\mu}&=& \frac{g_s^{-1}}{6!} H_7^{\mu a_1\cdots a_6} \mathfrak j_{6 a_1\cdots a_6}
+\frac{1}{2!} F_3^{\mu a_1 a_2} J_{2a_1a_2} 
\nonumber\\
&&+ \frac{3}{4!}H_3^{\mu a_1a_2} C_2^{a_3a_4}J_{4a_1\cdots a_4}~,
\\
\label{geneqae2}
&&d \star J_2 + H_3 \wedge \star J_4 =0~,
\\
\label{geneqae3}
&&d\star J_4 - \star \mathfrak j_6 \wedge F_3 =0~,
~~d\star \mathfrak j_6 =0
~.
\end{eqnarray}
\vskip -.1cm
Note that the index $\mu$ in \eqref{geneqae1} is an index in the ten-dimensional ambient KS background metric. Similarly, $\star$ in \eqref{geneqae2}, \eqref{geneqae3} is the Hodge dual for the 10d KS metric.

The differential equations \eqref{geneqae1}-\eqref{geneqae3} should be viewed as a system of dynamical equations for the unknown degrees of freedom of a 6d effective theory. These degrees of freedom include the transverse scalars (that express $\gamma_{ab}$), the functions $r_0,\alpha,\theta$ and the orthonormal vectors $u^a,v^a,w^a$.

The last equation in \eqref{geneqae3} is equivalent to the statement that the NS5-brane charge density $Q_5=\mathcal C r_0^2 \sinh\alpha \cosh\alpha \cos\theta$ is a constant of motion proportional to the number $N_5$ of NS5 branes. For fivebrane configurations wrapping an $S^2$ inside the $S^3$ in \eqref{geneqaa}, as considered below, the first equation in \eqref{geneqae3} similarly implies that the 3-brane Page charge $Q_3 = \int_{S^2} * (J_4 + \star(\star \mathfrak j_6 \wedge C_2))$ is also a constant of motion proportional to the induced D3-brane charge of the bound state. We denote $Q_3/(4\pi Q_5)= \pi p g_s \ell_s^2$, using $p$ from now on as the ratio of the number of anti-D3s to NS5s.

\paragraph{Recovering KPV at extremality.} 
Consider the extremal limit of the equations \eqref{geneqae1}-\eqref{geneqae3}. We focus on a possibly time-dependent configuration where the fivebrane bound state at $r=0$ wraps an $S^2$ inside the $S^3$ of the KS background. Writing the $S^3$ metric as $d\Omega_3^2= d\psi^2 + \sin^2\psi \left( d\omega^2 +\sin^2\omega d\varphi^2 \right)$ we turn on the single transverse scalar $\psi=\psi(t)$ in static gauge $(t=\sigma^0, x^i=\sigma^i, \omega=\sigma^4, \varphi=\sigma^5, ~ i=1,2,3)$ and set $v^a\partial_a = (\sqrt{g_s M}\ell_s b_0 \sin\psi)^{-1}\partial_\omega$, $w^a\partial_a = (\sqrt{g_s M}\ell_s b_0 \sin\psi \sin\omega)^{-1}\partial_\varphi$, $u^a\partial_a = (\sqrt{g_s M}\ell_s b_0 \sqrt{1-\psi'^2})^{-1} \partial_t$.
Then, the full set of equations \eqref{geneqae1}-\eqref{geneqae3} reduces to 
\vspace{-0.1cm}
\begin{eqnarray}
\label{KPVab}
\tan\theta = \frac{1}{b_0^2 \sin^2\psi} \left( \frac{\pi p}{M} - \psi + \frac{1}{2} \sin(2\psi) \right)
~,
\end{eqnarray}
\vspace{-0.3cm}
\begin{eqnarray}
\label{KPVac}
\begin{split}
\cot\psi &=& \frac{1}{b_0^2} \sqrt{1-\psi'^2}\sqrt{1+\tan^2\theta}+\frac{1}{b_0^2}\tan\theta 
\\
&&-\frac{1}{2}(1+\tan^2\theta) \frac{\psi''}{1-\psi'^2}
~.
\end{split}
\end{eqnarray}
After eliminating $\tan\theta$, it is trivial to check that the resulting equation for $\psi$ coincides with the Euler-Lagrange equations of the DBI action obtained in \cite{Kachru:2002gs} by S-duality of the D5 brane. In this manner, we recover the equations of motion of the KPV effective action from supergravity. A more general relation between the extremal blackfold equations and DBI equations can be derived along the lines of \cite{Niarchos:2015moa} (see also \cite{Grignani:2016bpq}).  

The maximum value $p^*$ that allows a metastable vacuum has the following meaning in the blackfold language. Since the NS5 branes at extremality have a nonvanishing Hagedorn local temperature $T_{\rm H}=\frac{1}{2\pi}\sqrt{\frac{\mathcal C}{Q_5}} \sqrt{|\cos\theta|}$, one can show that $p^*$ is very close to the point $\hat p$ where this temperature takes its maximum possible value. The latter
occurs when $\cos\hat \theta=1$, i.e.\ when the 3-brane charge is depleted. From Eq. \eqref{KPVac} we deduce that at that point $\hat \psi \simeq 0.7506$ obeying the equation $\cot\hat\psi = \frac{1}{b_0^2}$. Then, Eq.\ \eqref{KPVab} gives $\frac{\hat p}{M} = \hat \psi -\frac{1}{2}\sin(2\hat\psi) \simeq 0.08$, also noted in \cite{Kachru:2002gs}.

\begin{figure*}[!]
\begin{center}
\includegraphics[width=0.32\textwidth]{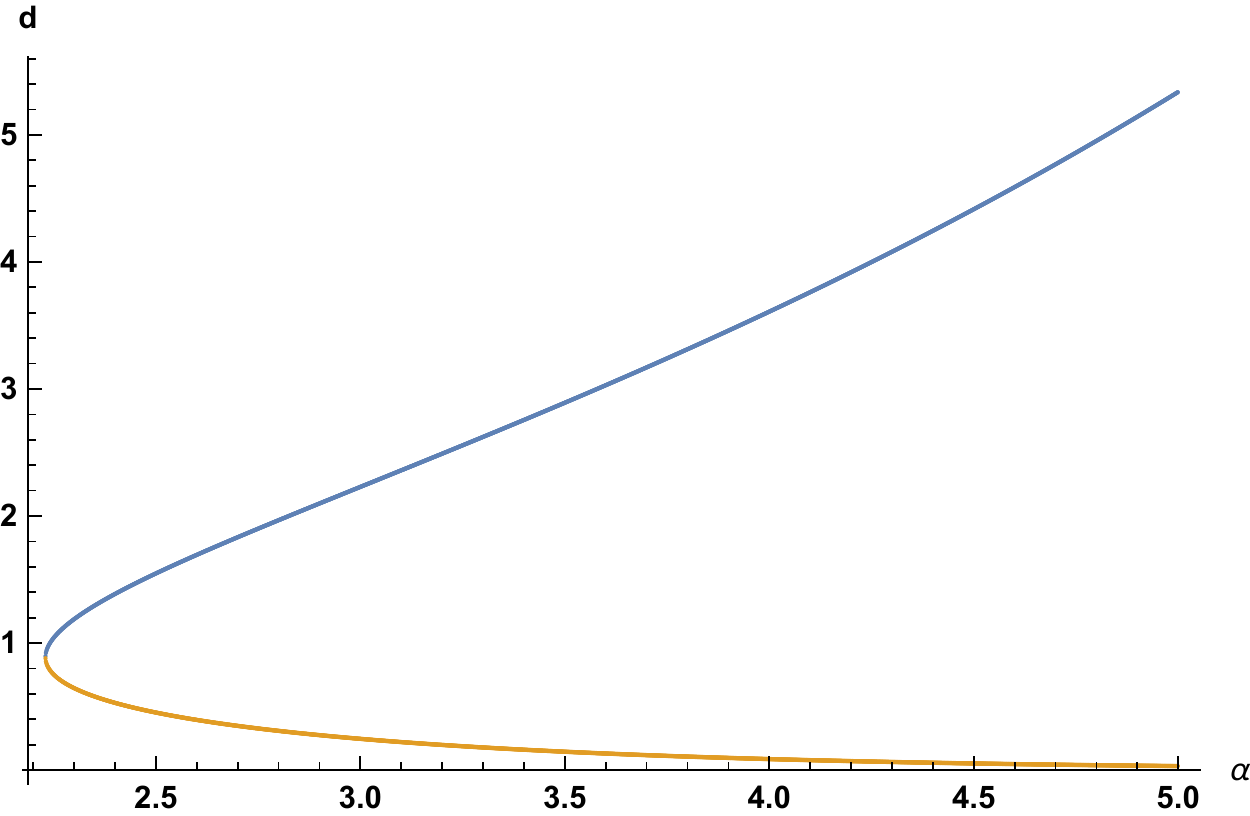} \hspace{2.5cm}
\includegraphics[width=0.34\textwidth]{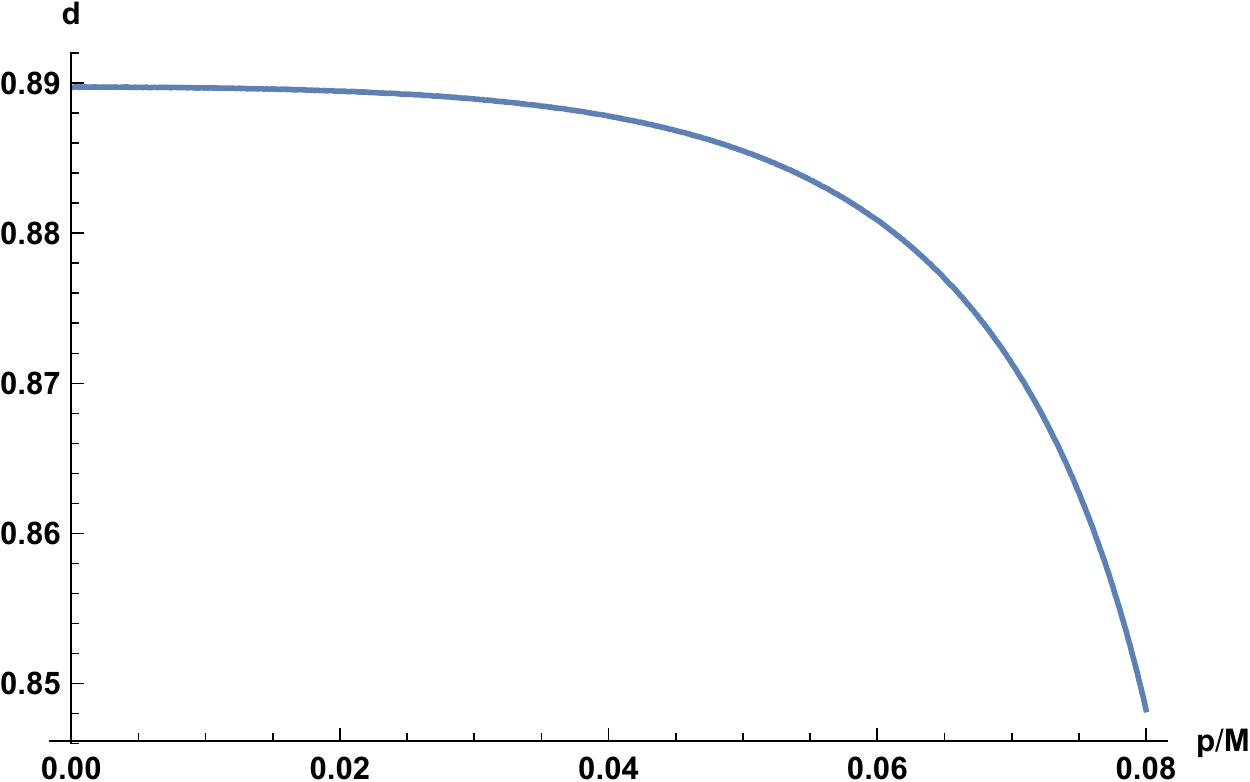} 
\end{center}
\vspace{-0.3cm}
\caption{Plots of the ratio $d$ that expresses how `fat' a non-extremal $\overline{\rm D3}$-NS5 bound state is. On the left plot we depict the dependence of $d$ on the non-extremality parameter $\alpha$ for the unstable (blue) and meta-stable (orange) branches for $p/M=0.03$. On the right plot, we depict $d$ at the critical merger point as a function of $p/M$.}
\vspace{-0.3cm}
\label{figd}
\end{figure*}

\vspace{0.2cm}
\paragraph{Non-extremal static configurations.}  

It is straightforward to repeat the above exercise for non-extremal configurations at finite $\alpha$. We continue to focus on the same ansatz for the worldvolume fields, now restricting to time-independent profiles. The non-extremal, static version of equation \eqref{KPVab} is exactly the same as before. On the other hand, when expressed in terms of $\alpha$  eq.\ \eqref{KPVac} becomes\begin{eqnarray}
\label{nonextab}
\cot\psi &=& \frac{1}{b_0^2}\left( \frac{\coth\alpha}{\cos\theta}+\tan\theta\right) \frac{2\cos^2\theta}{2\cos^2\theta+(\sinh\alpha)^{-2}}
.~~
\end{eqnarray}
This equation can be written alternatively in terms of the local temperature $T=\frac{1}{2\pi r_0 \cosh\alpha}$, or the local entropy density $s= 2\pi \mathcal C r_0^3 \cosh\alpha$, or any other quantity that characterises the deviation from extremality. We refer the reader to Sec.~I in Supplemental Material for a derivation of the equations of motion from an effective potential. In what follows, we will make use of an effective potential $V_S$ that keeps the entropy $S$ fixed. 

\paragraph{Results for non-extremal configurations.} 

Consider first the regime $p/M < p^*/M \simeq 0.08$, where the extremal solutions have a meta-stable vacuum. In Fig.\ \ref{fig1} we show how the effective  potential $V_S$ changes as we vary the entropy $S$ for a fixed value of $p/M$. We observe two interesting new features. Firstly, as soon as $S$ is turned on, a new unstable vacuum emerges (black dots on the right plot of Fig.\ \ref{fig1}) near the North pole, $\psi \simeq 0$. For sufficiently small values of $S$ there are three extrema: two unstable and one meta-stable. Secondly, as we increase $S$ further the new unstable extremum comes closer to the meta-stable vacuum and the two merge at a critical value of the entropy $S^*$, which is a function of $p/M$. Above this value the meta-stable vacuum is lost. The new unstable state represents a fat black NS5 with a highly pinched $\mathbb{R}^3\times S^5$ horizon geometry that resembles a black $\overline{\rm D3}$. Instead, the meta-stable state starts life near extremality as a thin black NS5 with $\mathbb{R}^3\times S^2 \times S^3$ horizon topology. At the merger the meta-stable black NS5 turns effectively into a black $\overline{\rm D3}$. The picture of a merger driven by horizon geometry is reinforced by the following observation. 

A quantitative measure of the `fatness' of a black NS5 wrapping an $S^2$ is provided by the ratio
$d = 2\sqrt{p} \hat r_0/(\sqrt{M}\sin(\psi))$ \cite{Note2}, where $\hat r_0 \equiv \sqrt{\mathcal C} r_0/\sqrt{Q_5}$ is dimensionless. The ratio $d$, which is a natural function of $p/M$ and the equilibrium $\psi$, compares the scale $2\sqrt{g_s p} \hat r_0 \ell_s= 2(g_s p N_5^{-1})^{1/2} r_0$ associated to the Schwarzschild radius and the scale of the $S^2$ wrapped by the NS5 worldvolume $\sqrt{g_s M}\ell_s \sin(\psi)$. As an illustration, on the left plot in Fig.\ \ref{figd} we see how $d$ behaves in the unstable branch (blue color) and the meta-stable branch (orange color). Recall that $\alpha\to \infty$ represents extremality. The unstable branch has visibly higher values of $d$, expressing the dominance of the Schwarzschild radius. The meta-stable branch captures a thin black NS5 with small values of $d$. The merger occurs at a value of $d$ notably close to 1. 

On the right plot of Fig.\ \ref{figd} we show how $d$ at the merger point behaves as a function of $p/M$. Remarkably, the ratio remains effectively constant, near the value $0.89$ over a significant range. It deviates slightly from this value in the vicinity of the upper bound of $p/M$, where effects from the second unstable state (already visible in KPV \cite{Kachru:2002gs}) become important. The characteristically weak dependence of $d$ on $p/M$ is a clear signal that the properties of the merger point are closely tied to the properties of the horizon geometry. 
Finally, by increasing $p/M$ further, above the critical value $p^*/M\simeq 0.08$, we observe the complete loss of the meta-stable vacuum exactly as in the extremal KPV analysis \cite{Kachru:2002gs}. The unstable vacuum in the vicinity of the North pole, however, remains, even above $p^*/M$, and constitutes the single vacuum of the non-extremal static blackfold equations.

\paragraph{Discussion.}
The results of the blackfold analysis are consistent in the regime of large $g_sp$ when  $N_5  \ll g_s M  \sin^2 \psi$ and $\sqrt{p/M} \ll \sqrt{g_s M} \sin^2\psi$ (see Sec.~II of Supplemental Material). For sufficiently large $M$ our calculations are therefore valid everywhere except for a small region around the North pole. At extremality we recover the results of KPV \cite{Kachru:2002gs} in a regime very different from theirs ($p\sim \mathcal{O}(1)$). Moving away from extremality was sofar impossible. Our analysis reveals two novel features of black $\overline{\rm D3}$-NS5 branes: a new unstable state near the North pole, and the persistence of the meta-stable state for small enough horizon radius. For a critical value of the horizon radius the meta-stable state and the new unstable state merge. Above this critical value, the meta-stable state is lost. Remarkably, these two features coincide with what was anticipated in  \cite{Cohen-Maldonado:2015ssa} based on a very complementary viewpoint that extracted features from would-be exact solutions of NS5 branes. A nogo-theorem, based on singularities, implies that the black NS5, if it exists, has a maximum horizon radius until it disappears. A black $\overline{\rm D3}$ also escapes the nogo, but was deemed unphysical in \cite{Cohen-Maldonado:2015ssa}, since it does not persist in the extremal limit. We have now shown that when the nogo's of \cite{Cohen-Maldonado:2015ssa, Cohen-Maldonado:2016cjh} do not apply we find go's. Together with the consistency of the extremal limit, this constitutes strong evidence for the existence of the meta-stable states, since they have now been argued for in complementary regimes.  Our analysis should impact the understanding of finite temperature effects in confining cascading gauge theories. We leave a study of this aspect to future work.

\begin{acknowledgments}
\vspace{0.2cm}

We thank F.~F.~Gautason and J.~P. van der Schaar for useful discussions.  The  work  of  TVR is  supported  by  the  FWO odysseus grant G.0.E52.14N and the C16/16/005 grant of the KULeuven.  We acknowledge support from the European Science Foundation HoloGrav Network. The work of V.N. is supported by STFC under the consolidated grant ST/P000371/1. The work of  NO is supported in part by a grant of the Independent
Research Fund Denmark (grant number DFF-6108-00340). JA is partly supported by the Netherlands Organisation for Scientific Research (NWO).
\end{acknowledgments}



\vspace{0.4cm}
\noindent
\begin{center}
{\bf Supplementary Material}
\end{center}

\vspace{-0.3cm}
\appendix

\section{Appendix I: Non-extremal thermodynamic potentials \label{app:pot}}  

Eqs.\ \eqref{KPVab}, \eqref{nonextab} can be obtained by extremising an appropriate  effective potential, which can be derived from thermodynamic considerations (see e.g.\ \cite{Emparan:2011hg, Armas:2018ibg}). Since the local temperature of the NS5 brane does not vanish at extremality, it is appropriate to choose an effective potential that holds some other quantity fixed, e.g.\ the global entropy $S=\int_{\mathcal B_5} \sqrt{-\gamma} s/\textbf{k}= 8\pi^2 (g_s M b_0^2)^{5/2} \mathcal C r_0^3 \cosh\alpha \sin^2\psi$, where $\mathcal B_5$ is the spatial part of the worldvolume $\mathcal M_6$, $s$ is the local entropy density defined above and ${\bf k}$ the Killing vector in the direction of the velocity vector $u$. Under the assumptions that the $\overline{\rm D3}$ is trivially embedded within the NS5, that the $\overline{\rm D3}$ directions are aligned with spatial background isometries, and of constant dilaton and vanishing 5-form field strength, the potential at fixed global entropy and charges $Q_3, Q_5$ is
\begin{equation}
\label{potaa}
V_S[\psi] = - \int_{\mathcal M_6}d^6 \sigma \sqrt{-\gamma} \varepsilon + Q_5 \int_{\mathcal M_6} {\mathbb P}[B_6]
~,
\end{equation}
with $\varepsilon = \frac{3}{2}\mathcal C r_0^2 + | Q_5 \sqrt{1+\tan^2\theta}| \tanh\alpha$ the local energy density (see \eqref{geneqac}) and ${\mathbb P}[B_6]$ the pullback of the background $B_6$-field \cite{Note1}. $V_S[\psi]$ is the total energy in the system and can be obtained from a Legendre transform of the Euclidean onshell action of the $\overline{\rm D3}$-NS5 bound state.

The potential \eqref{potaa} when Wick rotating along the time direction has the interpretation of an equilibrium partition function for a higher-form fluid. Rewriting the charge $Q_3$ as
\begin{equation}
Q_3=Q_5\int_{S^2}\left(\sqrt{\gamma_\perp}\tan\theta+\mathbb{P}^\perp[C_2]\right)~~,
\end{equation}
where $\gamma^\perp_{ab}$ is the metric on $S^2$ and $\mathbb{P}^\perp[C_2]$ the pullback of $C_2$ onto the $S^2$, direct variation with respect to the sources $\gamma_{ab}$, $C_{3}$ and $C_{6}$
yields the stress tensor \eqref{geneqac} and the currents $J_2$ and $\mathfrak j_6$ in \eqref{geneqad}, respectively \cite{Note3}. The global temperature of the system $T_{\text{H}}$ and chemical potentials $\Phi^{(3)}_{\text{H}}$ and $\Phi^{(5)}_{\text{H}}$ can be obtained from \eqref{potaa} according to
\begin{equation}
\begin{split}
&T_{\text{H}}=-\frac{1}{\sqrt{-\gamma}}\frac{\delta V_S}{\delta S}=\mathcal{T}\textbf{k}~,~\Phi^{(3)}_{\text{H}}=\frac{1}{\sqrt{-\gamma}}\frac{\delta V_S }{\delta Q_3}~~,\\
&\Phi^{(5)}_{\text{H}}=\frac{1}{\sqrt{-\gamma}}\frac{\delta V_S }{\delta Q_5}~~,
\end{split}
\end{equation}
and have the expected form that arises from a general analysis of higher-form fluids \cite{Armas:2018ibg}.

\section{Appendix II: Regimes of validity \label{app:val}}

Validity of the blackfold expansion requires a large separation of scales $r_b\ll \mathcal R, L$. In the case at hand, the characteristic length scale $r_b$ 
is the largest scale among the energy density radius ($r_\varepsilon\sim r_0\sinh\alpha$) and the scales associated to the NS5 and  $\overline{\rm D3}$ charge respectively.
The scale $\mathcal R$ is controlled by the size of the $S^2$ that the NS5 wraps, while the background scale $L$ is set by the size of the $S^3$. 
We also use  the fact that in all configurations of interest the two terms on the RHS of \eqref{KPVab} are either comparable or the $\pi p/M$ term dominates. For our purposes it is sufficient to consider  $r_h^{(NS5)} \ll \mathcal R$ and $r_h^{(\overline{\rm D3})}  \ll \mathcal R$ respectively leading to
\begin{equation}
\label{validaa}
\sqrt{\frac{N_5}{M}} \ll g_s \sin\psi~,~~ 
\sqrt{\frac{p}{M}}\ll g_s \sqrt{M} \sin^2\psi
.
\end{equation}
Both equations fail at the North pole, $\psi=0$. For sufficiently large $M$, however, our calculations will be valid everywhere except a small region around the North pole. In turn, the constraint $r_\varepsilon\ll \mathcal{R}$ leads to the requirement $d\ll g_s\sqrt{p}(\sqrt{N_5}\sinh\alpha)^{-1}$.

In addition, since the NS5 brane has a running dilaton one may worry whether
  regions of spacetime with large values of string coupling $e^\phi $
  invalidate our analysis. We note that the running of the dilaton is capped
  off at the horizon for non-extremal solutions at the value $e^\phi (r_0)= g_s
  \protect \sqrt {\protect \qopname \relax o{sin}^2\theta + \protect \qopname
  \relax o{cosh}^2\alpha \protect \qopname \relax o{cos}^2\theta }$. Hence, by
  suitably tuning the asymptotic value of $g_s$ we can achieve wide areas in
  parameter space where our solutions are everywhere weakly coupled.
  Admittedly, this tuning is not possible for extremal solutions. However,
  since it is understood how to treat the strong coupling singularity of NS5
  branes in flat space, and since the constraint (blackfold) equations can be
  obtained in a far-zone analysis of the solution, where the string coupling is
  weak, we anticipate that a large dilaton in the bulk of the solution does not
  invalidate the conclusions of our analysis even at extremality.

\renewcommand{\tt}{\normalfont\ttfamily}
\providecommand{\href}[2]{#2}\begingroup\raggedright\endgroup
\end{document}